\documentclass[draft]{agujournal2019}
\usepackage{url} 
\usepackage{lineno}
\usepackage[finalnew]{trackchanges} 
\usepackage{soul}

\draftfalse

\journalname{Geophysical Research Letters}

\begin{document}

%
%


\title{Sensitive Dependence of Global Climate to Continental Geometry}

%
%




\authors{Mark Baum\affil{1}, Minmin Fu\affil{1}, and Stephen Bourguet\affil{1}}


\affiliation{1}{Harvard University, 20 Oxford St., Cambridge MA 02138}





\correspondingauthor{Mark Baum}{markbaum@g.harvard.edu}




\begin{keypoints}
\item We simulate climate and weathering with an ensemble of randomly generated continental configurations.
\item Tropical land fraction and continent latitude are poor predictors of weathering rates, but degree of continental breakup is better.
\item Half of the weathering variability is produced by complex, hard to predict interaction between continent geometry and atmospheric dynamics.
\end{keypoints}

%
%

%
%


\begin{abstract} 
Over its multibillion-year history, the Earth has experienced a wide range of climates. The long-term climate is controlled by the atmospheric carbon dioxide concentration, which is regulated by marine sequestration through chemical weathering. This chemical weathering sink is strongly linked to the distribution and composition of the continents. However, the effect of continental distribution has never been studied within a general framework. Here we show that the global weathering rate is sensitive to the size and shape of the continents, but is not well explained by the amount of land in the tropics. We construct synthetic continental configurations and use an ensemble of global climate model simulations to isolate the expected effect of continental arrangement on weathering and carbon burial. Runoff patterns are complex, sensitive to detailed features of continental geometry, and poorly predicted by continental latitude. These results help explain the long-term variability and irregularity of Earth's climate.
\end{abstract}

\section*{Plain Language Summary}
Chemical weathering of the continental crust draws down atmospheric CO$_2$ and regulates the global climate on geological timescales. The weathering process is thought to be controlled primarily by temperature and runoff. Therefore, a concentration of continental landmasses in the tropics has long been considered a factor leading to higher weathering rates and global cooling. We rigorously test this hypothesis by running a large ensemble of climate simulations with random continental configurations. Surprisingly, we find that tropical land fraction and mean continental latitude are poor predictors of global weathering rates. We also find that although the size of the continents is important, it is not a dominant factor either. A significant fraction of weathering variability is driven by precipitation and runoff patterns, which are sensitive to detailed aspects of continental geometry. This sensitivity indicates that factors other than the latitudinal position of the continents (such as continental breakup or crustal composition) are important controls on the Earth's long-term climate.

%
%

%


%
%
%
%

\section{Introduction}
\label{sec:intro}

Geological and geochemical evidence indicates that Earth's climate has varied dramatically over at least the past 2.5 billion years. Three ``Snowball Earth" periods have been identified, when most or all of the planet's surface was covered in ice \cite{Hoffman-Schrag-2002:snowball, Pierrehumbert-Abbot-Voigt-et-al-2011:climate}. The first of these episodes occurred at the beginning of the Paleoproterozoic era, approximately 2.5~Gya \cite{Evans-Beukes-Kirschvink-1997:low, Kirschvink-Gaidos-Bertani-et-al-2000:paleoproterozoic}. The next two occurred during the Cryogenian period of the Neoproterozoic era, about 700~Mya and with inception times roughly 50~Ma apart \cite{Hoffman-Kaufman-Halverson-et-al-1998:neoproterozoic, Rooney-Strauss-Brandon-et-al-2015:cryogenian, Prave-Condon-Hoffmann-et-al-2016:duration}. At other points in time, like the Cretaceous and the early Eocene ``equable" climates, the Earth was warm and ice-free at high latitudes \cite{Berner-1990:atmospheric, Greenwood-Wing-1995:eocene}.

Why did Earth experience snowball climates at some points, but warm climates at others? Over geologic time, the atmospheric carbon dioxide (CO$_2$) concentration, and therefore the climate, is thought to be regulated by volcanic outgassing and the silicate weathering feedback \cite{Urey-1952:early, Walker-Hays-Kasting-1981:negative, Marshall-Walker-Kuhn-1988:long, Berner-Lasaga-1989:modeling}. In this picture, when atmospheric carbon dioxide levels change, the resulting change in temperature and runoff modifies the global cation flux to the ocean and the rate of marine carbonate burial, counteracting the initial CO$_2$ perturbation. Marine carbonate burial sequesters CO$_2$ from the ocean and atmosphere until it is recycled back by the subduction of oceanic crust \cite{Berner-Lasaga-Garrels-1983:carbonate}. Thus, atmospheric CO$_2$ has varied throughout Earth's history as rates of CO$_2$ input and burial have counterbalanced each other over millions of years \cite{Edmond-Huh-2003:non}.

Previous studies have identified changes in continental configuration as a possible driver of long-term climate change \cite{Schrag-Berner-Hoffman-et-al-2002:initiation, Hoffman-Schrag-2002:snowball, Godderis-Donnadieu-Nedelec-et-al-2003:sturtian, Donnadieu-Godderis-Ramstein-et-al-2004:snowball, Donnadieu-Godderis-Pierrehumbert-et-al-2006:geoclim, Rooney-Macdonald-Strauss-et-al-2014:re, Macdonald-Schmitz-Crowley-et-al-2010:calibrating, Godderis-Donnadieu-Carretier-et-al-2017:onset, Cox-Halverson-Stevenson-et-al-2016:continental, Macdonald-Swanson-Hysell-Park-et-al-2019:arc}. As the continents rift, drift, and collide, the changing pattern of temperature and precipitation over land, which controls the global weathering rate, shifts the atmospheric CO$_2$ concentration. Two aspects of the continental configuration are thought to affect the climate. The first is continental latitude. A high concentration of land masses in the tropics has long been considered a factor leading to global cooling \cite{Kirschvink-Gaidos-Bertani-et-al-2000:paleoproterozoic, Hoffman-Schrag-2002:snowball, Schrag-Berner-Hoffman-et-al-2002:initiation}. Low-latitude configurations raise the planetary albedo, strengthen weathering by exposing continental silicate to the warmest and wettest regions of the planet, and prevent weathering from being suppressed by ice growth at high latitudes. The second aspect is the size or, assuming fixed total land fraction, the number of continental land masses \cite{Donnadieu-Ramstein-Fluteau-et-al-2004:impact, Donnadieu-Godderis-Pierrehumbert-et-al-2006:geoclim, Godderis-Donnadieu-2019:sink}. On one end of this spectrum are supercontinents, which are expected to be quite dry in their interiors with limited weathering. On the other end are rifted configurations with many small land masses that should promote runoff and weathering.

This idea has been notably applied to the initiation of snowball climates. The first Cryogenian snowball, called the Sturtian glaciation, appears coincident with the breakup of the tropical supercontinent Rodinia \cite{Trindade-Macouin-2007:palaeolatitude, Li-Bogdanova-Collins-et-al-2008:assembly}. The increase in runoff following the breakup may have dramatically strengthened global weathering, thereby leading to dramatic cooling and glaciation \cite{Donnadieu-Godderis-Ramstein-et-al-2004:snowball}. However, there is no reason that the influence of continental configuration on the climate should be limited to specific episodes like the snowballs. If the influence is significant, it should be present throughout Earth's history.

Prior modeling studies of continental configuration and weathering do not distinguish between the effects of continental latitude and size or constrain their importance in a general manner. Simple arguments about the effect of continental latitude do not capture the influence of continental geometry on precipitation patterns, which are quite difficult to predict but may strongly influence weathering. However, when global climate models have been used to resolve precipitation patterns and simulate the effect of continental breakup, they have been restricted to a few continental configurations based on paleogeographic reconstructions. These reconstructions do not sample across independent ranges of tropical land fractions and continent sizes, confounding the effects of these factors. In addition, due to uncertainties in paleolongitude, the continental reconstructions used for these studies involve an element of subjectivity that may introduce bias.

\section{Simulating Random Continental Configurations}
\label{sec:sim}

We pursue a general approach to the question of how continental configuration influences climate, making as few assumptions about the relationship between continental geometry and weathering as possible. We create an ensemble of climate model simulations with randomly generated continental configurations, broadening the sample space and dissecting the influence of continental configuration on weathering more robustly. Using truncated, random spherical harmonic expansions, we construct groups of continental configurations that represent the breakup of a consolidated landmass, or ``supercontinent.''

Groups of spherical harmonic expansions are characterized by the relative weighting of high and low degrees. If low degrees are more heavily weighted, coarse structure dominates and the expansion resembles a consolidated supercontinent. If high degrees have more weight, fine structure is visible and the expansions resemble a rifted set of smaller continents. Different degrees are weighted according to a proportionality between the degree $d$ and the spectral power in the degree $S$
\begin{equation}
    S \propto d^p \, ,
    \label{eq:sdp}
\end{equation}
where $p$ defines the relative weight of different degrees and can be thought of as a ``continental consolidation parameter." Supercontinents are generated by $p=-3$, which suppresses high-frequency harmonics. Intermediate consolidation is produced by $p=-2$ and rifted configurations by $p=-1$. Further details are provided in the Supporting Information.

\add{Importantly, we do not simulate the effects of large-scale topography, which promotes runoff over steeper land surfaces and can influence precipitation patterns via orographic rainfall. Indeed, recent studies have identified the importance of tropical arc-continents, which are thought to drive cooling by exhuming fresh ultramafic rock with steep topography.} \cite{Macdonald-Swanson-Hysell-Park-et-al-2019:arc, park2020emergence} \add{These are important considerations that will make for interesting future research and we return them in the Discussion. However, this study is focused on isolating the effects large-scale continental shape and arrangement. As such, like in previous studies on this topic, we simulate nearly flat land masses.}

Using 120 independent continental configurations (40 for each $p$ value), all with global land fractions of 30~\%, we simulate the mean temperature and runoff fields using the Community Earth System Model (CESM) at a resolution of $\sim$4$^{\circ}$ per cell. We use Neoproterozoic solar insolation of 1285~W/m$^2$, no vegetation, and an atmospheric CO$_2$ concentration of 1000~ppm. Obliquity is set to 23.5{\textdegree} and eccentricity is set to zero. This configuration produces mean tropical temperatures close to 295~K. Except for the continental configuration, simulations are identical.

\begin{figure}
    \centering
    \includegraphics[width=\textwidth]{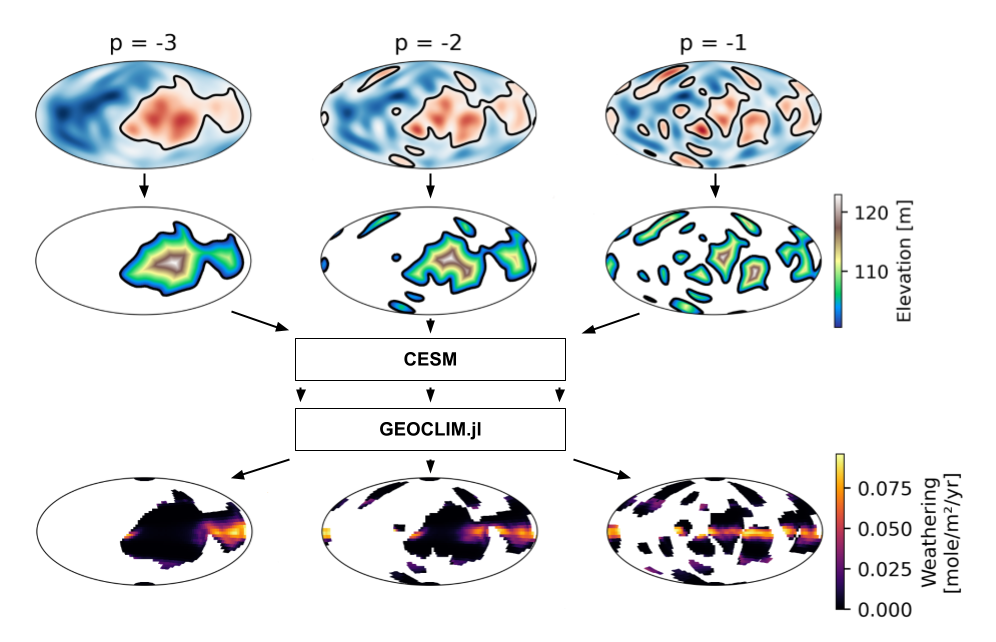}
    \caption{A diagram of the simulation steps for an example group of continental configurations. The upper row shows three truncated spherical harmonic expansions with different $p$ values, starting from the same random expansion coefficients and with the land (red) fraction set to 30~\%. The second row shows the same expansions converted to nearly flat topography for CESM. Simulated temperature and runoff fields are used to estimate weathering and MAC weathering rates are shown in the bottom row, computed using our publicly available weathering software module named \texttt{GEOCLIM.jl} \cite{baum_fu_geoclim}.}
    \label{fig:topo_321}
\end{figure}

For each simulated temperature and runoff distribution, we estimate global weathering rates using two formulations. The first formulation, known as the WHAK model \cite{Walker-Hays-Kasting-1981:negative}, includes the conventional exponential dependence on temperature. The second, known as the MAC model \cite{Maher-Chamberlain-2014:hydrologic}, includes a thermodynamic limit on the weathering rate and is generally less sensitive to differences in temperature than the WHAK model. The equations and parameters defining these models are given in the Supporting Information and our weathering code is publicly available \cite{baum_fu_geoclim}. We apply both of these models to all cells of the simulated climatologies and perform area-weighted sums to estimate global weathering. Figure \ref{fig:topo_321} diagrams the simulation process for one group of continental configurations, from spherical harmonic expansions to weathering rates.

\section{Results}
\label{sec:results}

\subsection{Continent Latitude}

\begin{figure}[ht!]
    \centering
    \includegraphics[width=0.6\textwidth]{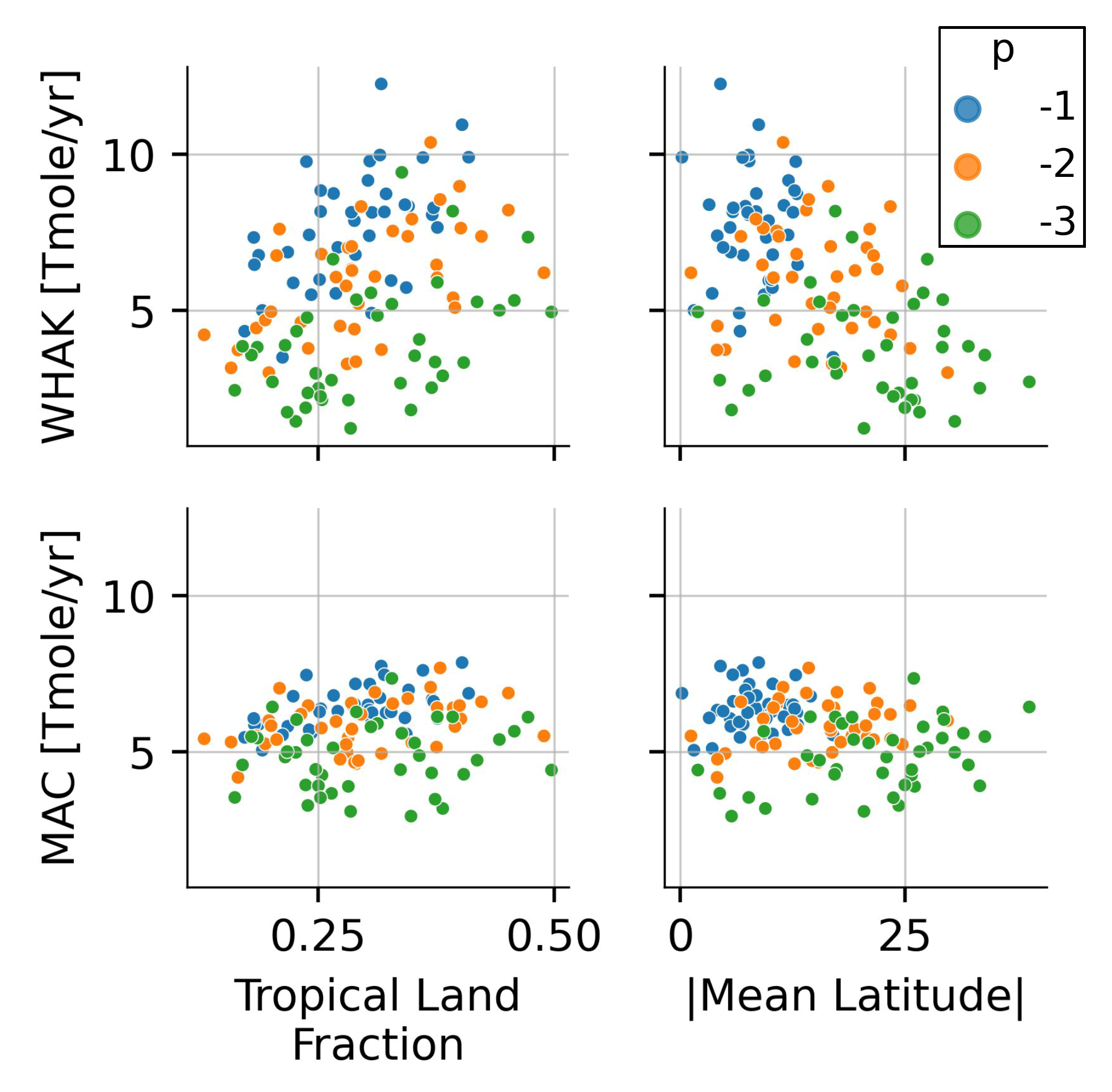}
    \caption{Scatter plots of TLF and absolute mean continental latitude against global WHAK and MAC weathering rates, grouped by the consolidation parameter $p$. Supercontinents correspond to ${p=-3}$ and are represented by green dots. Broken-up configurations correspond to ${p=-1}$, and are represented by blue dots. \change{}{TLF for modern day is approximately 23\%.}}
    \label{fig:scatter_composite}
\end{figure}

The left two panels of Figure \ref{fig:scatter_composite} show scatter plots of the estimated global weathering rate for all ensemble members against tropical land fraction (TLF), defined here as the fraction of land area within 15$^{\circ}$ of the equator. Results are not sensitive to this particular latitude cutoff. Points are colored according to the consolidation parameter $p$. If TLF is a good predictor of the global weathering rate, a strong positive correlation between these quantities should appear. However, there is only a weak correlation between TLF and WHAK weathering, with scarcely any discernible relationship between TLF and MAC weathering. Tropical land fraction, at least between 0.2 and 0.5, appears to be a poor predictor of global weathering rates.

The relationship between global weathering and continental latitude could be expressed in a number of ways and TLF is only one metric. The right panels of Figure \ref{fig:scatter_composite} show a second metric, scattering the absolute mean latitude of the landmasses against estimated weathering. The mean latitude is the area-weighted average of continental latitude for each configuration, analogous to the center of mass. If continental latitude is a primary factor governing weathering, we expect a clear negative correlation. Instead, we find almost no correlation for MAC weathering and, again, only a weak relationship for WHAK weathering. As the top right panel of Figure \ref{fig:scatter_composite} shows, the highest weathering rates are generally produced by configurations with low mean latitude. However, as we discuss later, this is primarily due to the consolidation parameter $p$. Within each $p$ group, there is no clear relationship. Mean continental latitude, at least when less than about 40$^{\circ}$, does not reliably predict global weathering.

Because landmass is more concentrated for supercontinents ($p=-3$), the range of mean latitudes for this group is wider and the effect of latitude should be most apparent. If land latitude is an important control on weathering, shifting supercontinental landmass from moderate latitude into the tropics should reliably increase weathering. Figure \ref{fig:scatter_composite} shows that it does not. For WHAK and MAC, the weathering rates produced by $p=-3$ members of our ensemble have no clear relationship with continental latitude. \add{The MAC model, in particular, is more sensitive to changes in runoff than temperature over most Earth-like parameters (see Supporting Information). Therefore, in the MAC formulation, landmasses in the tropics are not expected to weather more due to warmth alone, but only if they receive higher levels of precipitation and runoff. This strong runoff-dependence helps explain the particularly weak correlation between MAC weathering and TLF/mean latitude} \ref{fig:scatter_composite}.

\subsection{Continent \change{Size}{Fragmentation}}

\begin{figure}
    \centering
    \includegraphics[width=0.45\textwidth]{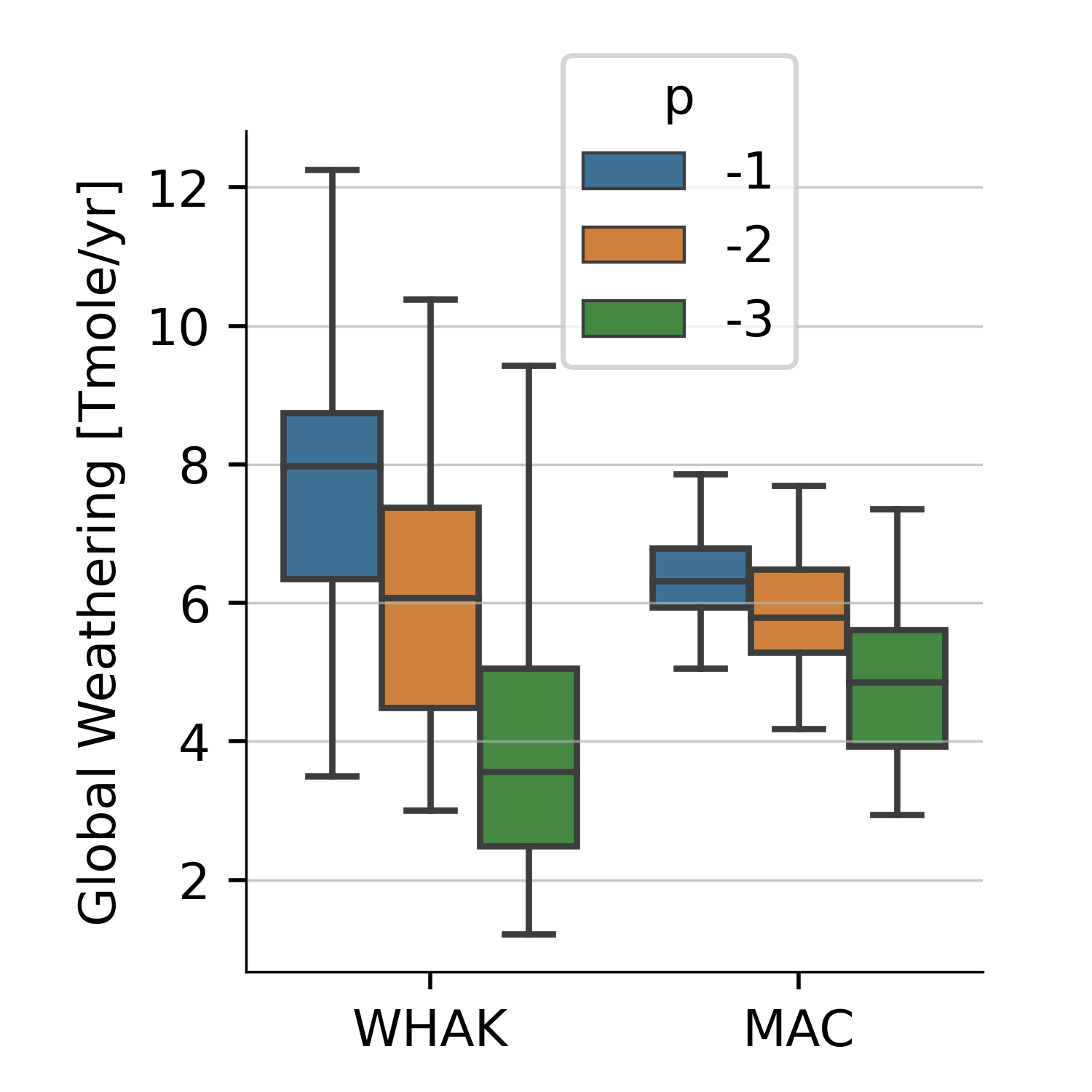}
    \caption{A box and whisker plot showing the effect of continental consolidation on global weathering estimates for the ensemble. Boxes indicate the interquartile range of each group, with the range indicated by the whiskers and the median indicated by the line in each box. The blue ($p=-1$) boxes represent distributed configurations and the green boxes ($p=-3$) represent consolidated/supercontinent configurations. Each box represents 40 ensemble members. Continental breakup has a significant effect on weathering, amounting to an average increase of about 100~\% for WHAK weathering and \change{50~\%}{35~\%} for MAC weathering.}
    \label{fig:p_boxes}
\end{figure}

We also examine the effect of continental consolidation on global weathering rates. Figure \ref{fig:p_boxes} shows a summary for the two weathering formulae, again split by the consolidation parameter $p$. On average, continental consolidation has a clear effect on global weathering rates. WHAK weathering is much more sensitive, with larger ranges within each $p$ group and bigger shifts between the groups. A shift from $p=-3$ to $p=-1$, representing the breakup of a supercontinent, increases WHAK weathering by more than a factor of two, on average. MAC weathering is less sensitive, but the effect of continental consolidation is still notable. In this case, complete supercontinent breakup produces an average weathering increase of about 35~\%.

\subsection{Statistical Relationships}

Our results indicate that continental latitude is not a reliable predictor of global weathering but continental consolidation has a more significant influence. To evaluate these relationships more concretely, we standardize the results and construct linear regressions. The tropical land fraction, alone, achieves a coefficient of determination ($r^2$) of 0.11 for WHAK and 0.03 for MAC. These results are similar for regressions with mean continental latitude. Continental latitude metrics explain only a small portion of the variation in global weathering across the ensemble. Within groups with identical $p$ values, the highest $r^2$ value is 0.32, occurring in the regression between TLF and WHAK weathering with $p=-2$.

Regression with the consolidation parameter $p$ as the sole predictor achieves $r^2=0.41$ for WHAK and $r^2=0.36$ for MAC. These results are very similar whether $p$ is treated as a continuous variable or a categorical variable where each $p$ value is represented by an individual binary variable. \change{Finally,}{When we incorporate both tropical land fraction and the consolidation parameter ($p$) as predictors, the resulting model performs better, with respect to the $r^2$ metric, than with either predictor in isolation.} Multiple regression with TLF and $p$ yields $r^2$ values of 0.56 and 0.41 for WHAK and MAC, respectively. Continental latitude and consolidation, combined, explain about half of the weathering variability in the ensemble. A summary of the regression results is shown in Table \ref{tab:r2}.

\begin{table}
    \centering
    \begin{tabular}{ c | c | c }
        Predictor(s) & WHAK $r^2$ & MAC $r^2$ \\
        \hline
        Tropical Land Fraction (TLF) & 0.11 & 0.03 \\
        $|$Mean Latitude$|$ & 0.08 & 0.03 \\
        Consolidation Parameter ($p$) & 0.41 & 0.32 \\
        TLF \& $p$ & 0.56 & 0.41
    \end{tabular}
    \vspace{0.3cm}
    \caption{Summary of the coefficients of determination ($r^2$) for linear regressions where different continental characteristics are used to predict global mean weathering rates. The first three rows show the results of regression with an individual predictor. The last column shows the result of multiple regression using TLF and $p$.}
    \label{tab:r2}
\end{table}

\section{Sensitivity to Continental Geometry}

Our simulations isolate the effect of continental geometry on global weathering rates. Tropical land fraction and continental latitude are poor predictors of global weathering, at least within our simulated ranges. Continental consolidation is more significant, on average, but there is large variability within groups of simulations with identical $p$ values. The consolidation state of the continents, alone, is not a dominant control on global weathering. Together, continental latitude and consolidation capture about half of the global weathering variability in our ensemble.

\begin{figure}
    \centering
    \includegraphics[width=0.4\textwidth]{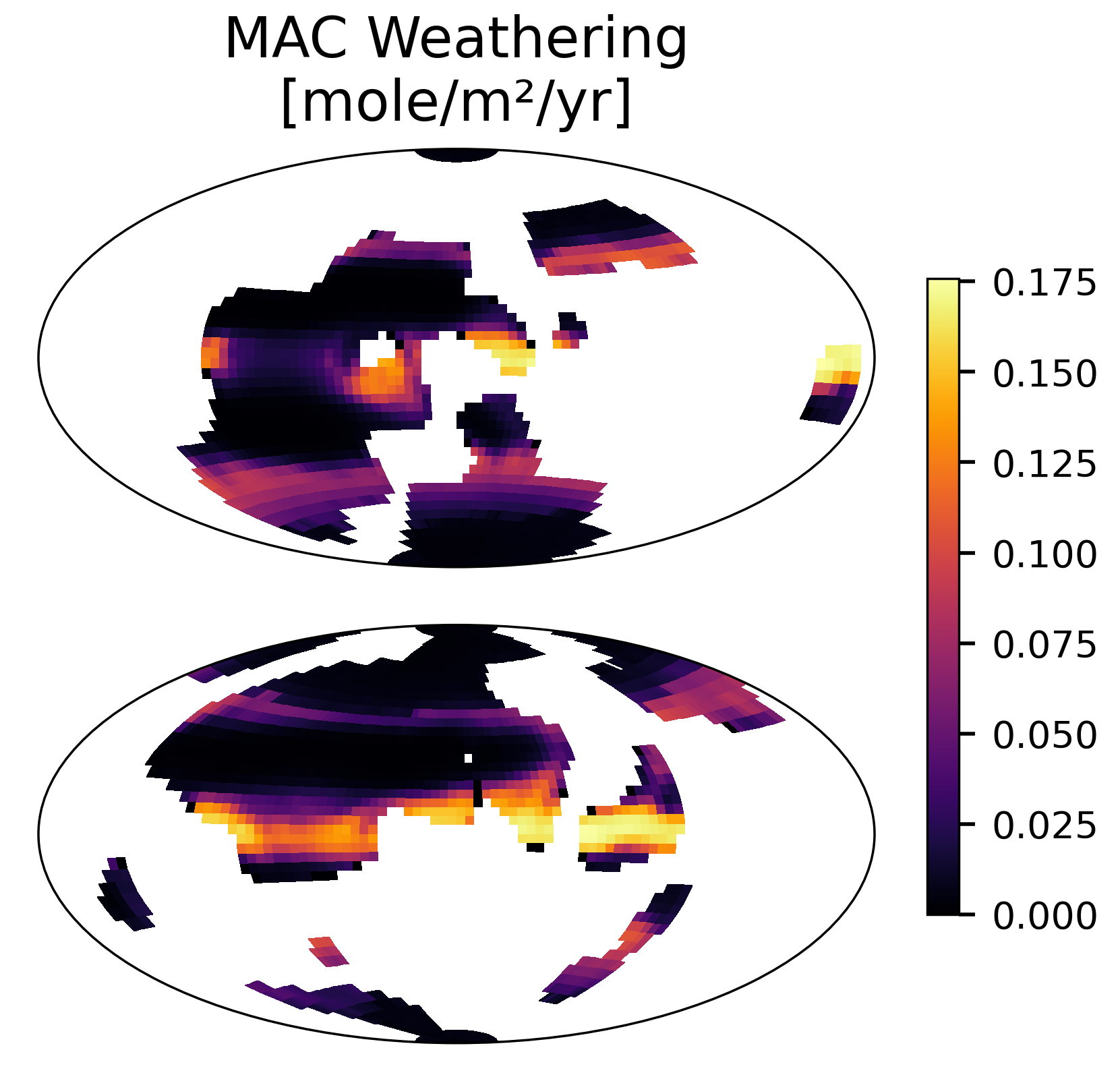}
    \caption{The spatial distribution of global MAC weathering for two configurations with $p=-2$ and nearly identical tropical land fractions but quite different global weathering rates. The top and bottom configurations have nearly equal TLFs of 29 and 29.5~\%, respectively. Global MAC weathering is 1.3 times higher in the bottom configuration, however, and global WHAK weathering is 2.5 times higher, owing to the variability in precipitation patterns induced by continental geometry.}
    \label{fig:w_diptych}
\end{figure}

What else explains the other half of the variability in our global weathering estimates? Configurations with similar latitude metrics and identical $p$ values can produce very different weathering rates. Figure \ref{fig:scatter_composite} shows this in general and Figure \ref{fig:w_diptych} illustrates a specific example. Because all simulations are performed with identical atmospheric CO$_2$ concentrations, they exhibit very similar temperatures, especially over the tropical ocean and coasts. Temperature differences across simulations do not explain weathering differences. The primary difference between simulations is the precipitation pattern over land. This pattern drives runoff and is strongly influenced by the continental configuration. The complex interaction between continental geometry and atmospheric dynamics has a strong influence on global weathering, responsible for the other half of our simulated weathering variability.

Sensitivity to precise features of continental geometry may help explain the lack of regular snowball episodes throughout Earth's history and the long term irregularity of the global climate record more generally. Assembly and rifting of the supercontinents Pangea and Rodinia have been associated with contemporaneous changes in climate \cite{Donnadieu-Godderis-Ramstein-et-al-2004:snowball, Donnadieu-Godderis-Pierrehumbert-et-al-2006:geoclim, brune2017potential}. However, no such change occurred during the warm and stable climate of the ``Boring Billion," 1.8-0.8~Gya, even though paleomagnetic evidence indicates that a low-latitude supercontinent rifted between 1.5-1.2~Gya \cite{Evans-2013:reconstructing}. Differences in the detailed geometry of these configurations may have produced quite different weathering and climate responses. More generally, the variable interaction between continental geometry and weathering could help explain the somewhat unpredictable nature of Earth's long-term climate history, especially considering additional sensitivity to continental composition \cite{Macdonald-Swanson-Hysell-Park-et-al-2019:arc}.

An important factor in global precipitation patterns that we do not address is topography. \add{As mentioned in the Methods section, topography may introduce another significant layer of weathering varibility.} In the modern climate, topography strongly influences monsoons and orographic rainfall. \add{It is also thought to be a primary factor in the substantial modern weathering contribution of the Southeast Asian islands} \cite{park2020emergence}. \add{Because we prioritized a large sample of random configurations, computational limits required us to simulate climatologies at about $\sim$4$^{\circ}$ per cell. This is a reasonable resolution and an improvement over many prior studies. However, it is too coarse to resolve small land masses like the Southeast Asian islands and is, in general, too coarse to capture the effects of orography and slope on weathering.}

Generating realistic, synthetic topography for random continental arrangements would be a challenge \change{and may introduce considerable subjectivity}{, but may be possible by incorporating assumptions regarding the tectonic setting of simulated landmasses. This is especially plausible for simulations involving reconstructions of Earth's real configuration history}. We have assumed that topography would not systematically change our results, as its effect on rainfall patterns is also complex and variable, but this should be a primary subject of future studies\remove{. Future work could also attempt to reproduce our results with high-resolution climate simulations, including topographic effects or not}, although it would be very computationally expensive to reproduce \change{the whole}{a similarly large} ensemble at much higher resolution.

Future work could attempt to identify precisely what geometric features of continental configurations promote or suppress weathering through their influence on precipitation \change{, but this would likely require an expanded ensemble}{and this may be a rich area of investigation}. This research could overlap and draw inspiration from efforts to understand future rainfall trends. With a large enough ensemble, machine learning techniques like convolutional neural networks could replace the GCM and weathering calculations. Such an emulator might help identify general patterns in continental arrangements that promote or suppress weathering and are difficult to otherwise detect.

\add{In conclusion, we have simulated the effects of large-scale continental configuration on estimates of global weathering rates with a random ensemble of continental arrangements. The size of our ensemble and the generality of these configurations enables us to examine metrics like continental latitude and size/number more robustly than was previously possible. In the broadest terms, latitude does not appear to a good predictor of global weathering. However, some tropical land mass is probably required and our ensemble does not represent the effects of relatively small, topographically important features like arc-continents. Continental size is important, but the complexity of rainfall patterns over the continents drives considerable variability in weathering. Further work will hopefully unravel this complexity, its impact on climate, and the role of topography and lithology.}

\section{Open Research}
Weathering calculations are performed with \texttt{GEOCLIM.jl} \cite{baum_fu_geoclim}, our publicly available weathering module written in the Julia language \cite{Bezanson2012, bezanson2017julia}. Statistical modeling is performed with the publicly available Python package \texttt{statsmodels} \cite{seabold2010statsmodels}. Plots and figures were created using the Python package \texttt{matplotlib} \cite{Hunter:2007}. The source code for CESM1.2 is freely available at \url{http://www.cesm.ucar.edu/models/cesm1.2/} following registration. All project files and results that are not otherwise part of publicly available software packages are permanently archived and freely available through Zenodo \cite{mark_baum_2022_6540321}.

\acknowledgments
We thank Dorian Abbot and R. J. Graham for introducing us to the MAC weathering formulation. We thank Erik Kluzek for technical assistance in the DiscussCESM online forum. We also thank Eli Tziperman for support during the course of this project and Katherine Keller for helpful discussions. M.F. was supported by NSF Climate Dynamics program (joint NSF/NERC) grant AGS-1924538. High-performance computing resources on Cheyenne was provided by NCAR's Computational and Information Systems Laboratory, sponsored by the National Science Foundation (doi:10.5065/D6RX99HX).



\nocite{Abbot-2016:analytical}
\nocite{Alekseyev-Medvedeva-Prisyagina-et-al-1997:change}
\nocite{Gerlach-2011:volcanic}
\nocite{Graham-Pierrehumbert-2020:thermodynamic}
\nocite{Haqq-Misra-Kopparapu-Batalha-et-al-2016:limit}
\nocite{Lagache-1976:new}
\nocite{Lawrence-Oleson-Flanner-et-al-2011:parameterization}
\nocite{Maher-Steefel-DePaolo-et-al-2006:mineral}
\nocite{Maher-Steefel-White-et-al-2009:role}
\nocite{Manabe-Wetherald-1975:effects}
\nocite{Neale-Chen-Gettelman-et-al-2012:description}
\nocite{Winnick-Maher-2018:relationships}
\bibliography{export}

\begin{thebibliography}{}

\bibitem [\protect \citeauthoryear {%
Abbot%
}{%
Abbot%
}{%
{\protect \APACyear {2016}}%
}]{%
Abbot-2016:analytical}
\APACinsertmetastar {%
Abbot-2016:analytical}%
\begin{APACrefauthors}%
Abbot, D\BPBI S.%
\end{APACrefauthors}%
\unskip\
\newblock
\APACrefYearMonthDay{2016}{}{}.
\newblock
{\BBOQ}\APACrefatitle {Analytical investigation of the decrease in the size of
  the habitable zone due to a limited CO2 outgassing rate} {Analytical
  investigation of the decrease in the size of the habitable zone due to a
  limited co2 outgassing rate}.{\BBCQ}
\newblock
\APACjournalVolNumPages{The Astrophysical Journal}{827}{2}{117}.
\PrintBackRefs{\CurrentBib}

\bibitem [\protect \citeauthoryear {%
Alekseyev%
, Medvedeva%
, Prisyagina%
, Meshalkin%
\BCBL {}\ \BBA {} Balabin%
}{%
Alekseyev%
\ \protect \BOthers {.}}{%
{\protect \APACyear {1997}}%
}]{%
Alekseyev-Medvedeva-Prisyagina-et-al-1997:change}
\APACinsertmetastar {%
Alekseyev-Medvedeva-Prisyagina-et-al-1997:change}%
\begin{APACrefauthors}%
Alekseyev, V\BPBI A.%
, Medvedeva, L\BPBI S.%
, Prisyagina, N\BPBI I.%
, Meshalkin, S\BPBI S.%
\BCBL {}\ \BBA {} Balabin, A\BPBI I.%
\end{APACrefauthors}%
\unskip\
\newblock
\APACrefYearMonthDay{1997}{}{}.
\newblock
{\BBOQ}\APACrefatitle {Change in the dissolution rates of alkali feldspars as a
  result of secondary mineral precipitation and approach to equilibrium}
  {Change in the dissolution rates of alkali feldspars as a result of secondary
  mineral precipitation and approach to equilibrium}.{\BBCQ}
\newblock
\APACjournalVolNumPages{Geochimica et Cosmochimica Acta}{61}{6}{1125--1142}.
\PrintBackRefs{\CurrentBib}

\bibitem [\protect \citeauthoryear {%
Baum%
\ \BBA {} Fu%
}{%
Baum%
\ \BBA {} Fu%
}{%
{\protect \APACyear {2022}}%
{\protect \APACexlab {{\protect \BCnt {1}}}}}]{%
mark_baum_2022_6540321}
\APACinsertmetastar {%
mark_baum_2022_6540321}%
\begin{APACrefauthors}%
Baum, M.%
\BCBT {}\ \BBA {} Fu, M.%
\end{APACrefauthors}%
\unskip\
\newblock
\APACrefYearMonthDay{2022{\protect \BCnt {1}}}{{\APACmonth{02}}}{}.
\newblock
\APACrefbtitle {Ensemble Weathering Results \& Project Files.} {Ensemble
  weathering results \& project files.}
\newblock
\APACaddressPublisher{}{Zenodo}.
\newblock
\begin{APACrefURL} \url{https://doi.org/10.5281/zenodo.6540321}
  \end{APACrefURL}
\newblock
\begin{APACrefDOI} \doi{10.5281/zenodo.6540321} \end{APACrefDOI}
\PrintBackRefs{\CurrentBib}

\bibitem [\protect \citeauthoryear {%
Baum%
\ \BBA {} Fu%
}{%
Baum%
\ \BBA {} Fu%
}{%
{\protect \APACyear {2022}}%
{\protect \APACexlab {{\protect \BCnt {2}}}}}]{%
baum_fu_geoclim}
\APACinsertmetastar {%
baum_fu_geoclim}%
\begin{APACrefauthors}%
Baum, M.%
\BCBT {}\ \BBA {} Fu, M.%
\end{APACrefauthors}%
\unskip\
\newblock
\APACrefYearMonthDay{2022{\protect \BCnt {2}}}{{\APACmonth{05}}}{}.
\newblock
\APACrefbtitle {markmbaum/GEOCLIM.jl: v0.1.12.} {markmbaum/geoclim.jl:
  v0.1.12.}
\newblock
\APACaddressPublisher{}{Zenodo}.
\newblock
\begin{APACrefURL} \url{https://doi.org/10.5281/zenodo.6533814}
  \end{APACrefURL}
\newblock
\begin{APACrefDOI} \doi{10.5281/zenodo.6533814} \end{APACrefDOI}
\PrintBackRefs{\CurrentBib}

\bibitem [\protect \citeauthoryear {%
Berner%
}{%
Berner%
}{%
{\protect \APACyear {1990}}%
}]{%
Berner-1990:atmospheric}
\APACinsertmetastar {%
Berner-1990:atmospheric}%
\begin{APACrefauthors}%
Berner, R\BPBI A.%
\end{APACrefauthors}%
\unskip\
\newblock
\APACrefYearMonthDay{1990}{}{}.
\newblock
{\BBOQ}\APACrefatitle {Atmospheric carbon dioxide levels over {Phanerozoic}
  time} {Atmospheric carbon dioxide levels over {Phanerozoic} time}.{\BBCQ}
\newblock
\APACjournalVolNumPages{Science}{249}{4975}{1382--1386}.
\PrintBackRefs{\CurrentBib}

\bibitem [\protect \citeauthoryear {%
Berner%
\ \BBA {} Lasaga%
}{%
Berner%
\ \BBA {} Lasaga%
}{%
{\protect \APACyear {1989}}%
}]{%
Berner-Lasaga-1989:modeling}
\APACinsertmetastar {%
Berner-Lasaga-1989:modeling}%
\begin{APACrefauthors}%
Berner, R\BPBI A.%
\BCBT {}\ \BBA {} Lasaga, A\BPBI C.%
\end{APACrefauthors}%
\unskip\
\newblock
\APACrefYearMonthDay{1989}{03}{}.
\newblock
{\BBOQ}\APACrefatitle {Modeling the Geochemical Carbon Cycle} {Modeling the
  geochemical carbon cycle}.{\BBCQ}
\newblock
\APACjournalVolNumPages{Scientific American}{222}{}{74-82}.
\newblock
\begin{APACrefDOI} \doi{10.1038/scientificamerican0389-74} \end{APACrefDOI}
\PrintBackRefs{\CurrentBib}

\bibitem [\protect \citeauthoryear {%
Berner%
, Lasaga%
\BCBL {}\ \BBA {} Garrels%
}{%
Berner%
\ \protect \BOthers {.}}{%
{\protect \APACyear {1983}}%
}]{%
Berner-Lasaga-Garrels-1983:carbonate}
\APACinsertmetastar {%
Berner-Lasaga-Garrels-1983:carbonate}%
\begin{APACrefauthors}%
Berner, R\BPBI A.%
, Lasaga, A\BPBI C.%
\BCBL {}\ \BBA {} Garrels, R\BPBI M.%
\end{APACrefauthors}%
\unskip\
\newblock
\APACrefYearMonthDay{1983}{}{}.
\newblock
{\BBOQ}\APACrefatitle {The carbonate-silicate geochemical cycle and its effect
  on atmospheric carbon dioxide over the past 100 million years} {The
  carbonate-silicate geochemical cycle and its effect on atmospheric carbon
  dioxide over the past 100 million years}.{\BBCQ}
\newblock
\APACjournalVolNumPages{American Journal of Science}{283}{7}{641--683}.
\newblock
\begin{APACrefDOI} \doi{10.2475/ajs.283.7.641} \end{APACrefDOI}
\PrintBackRefs{\CurrentBib}

\bibitem [\protect \citeauthoryear {%
Bezanson%
, Edelman%
, Karpinski%
\BCBL {}\ \BBA {} Shah%
}{%
Bezanson%
\ \protect \BOthers {.}}{%
{\protect \APACyear {2017}}%
}]{%
bezanson2017julia}
\APACinsertmetastar {%
bezanson2017julia}%
\begin{APACrefauthors}%
Bezanson, J.%
, Edelman, A.%
, Karpinski, S.%
\BCBL {}\ \BBA {} Shah, V\BPBI B.%
\end{APACrefauthors}%
\unskip\
\newblock
\APACrefYearMonthDay{2017}{}{}.
\newblock
{\BBOQ}\APACrefatitle {Julia: {A} fresh approach to numerical computing}
  {Julia: {A} fresh approach to numerical computing}.{\BBCQ}
\newblock
\APACjournalVolNumPages{SIAM review}{59}{1}{65--98}.
\newblock
\begin{APACrefURL} \url{https://doi.org/10.1137/141000671} \end{APACrefURL}
\PrintBackRefs{\CurrentBib}

\bibitem [\protect \citeauthoryear {%
Bezanson%
, Karpinski%
, Shah%
\BCBL {}\ \BBA {} Edelman%
}{%
Bezanson%
\ \protect \BOthers {.}}{%
{\protect \APACyear {2012}}%
}]{%
Bezanson2012}
\APACinsertmetastar {%
Bezanson2012}%
\begin{APACrefauthors}%
Bezanson, J.%
, Karpinski, S.%
, Shah, V\BPBI B.%
\BCBL {}\ \BBA {} Edelman, A.%
\end{APACrefauthors}%
\unskip\
\newblock
\APACrefYearMonthDay{2012}{}{}.
\newblock
{\BBOQ}\APACrefatitle {Julia: {A} Fast Dynamic Language for Technical
  Computing} {Julia: {A} fast dynamic language for technical computing}.{\BBCQ}
\newblock
\APACjournalVolNumPages{CoRR}{abs/1209.5145}{}{}.
\newblock
\begin{APACrefURL} \url{http://arxiv.org/abs/1209.5145} \end{APACrefURL}
\PrintBackRefs{\CurrentBib}

\bibitem [\protect \citeauthoryear {%
Brune%
, Williams%
\BCBL {}\ \BBA {} Mueller%
}{%
Brune%
\ \protect \BOthers {.}}{%
{\protect \APACyear {2017}}%
}]{%
brune2017potential}
\APACinsertmetastar {%
brune2017potential}%
\begin{APACrefauthors}%
Brune, S.%
, Williams, S\BPBI E.%
\BCBL {}\ \BBA {} Mueller, R\BPBI D.%
\end{APACrefauthors}%
\unskip\
\newblock
\APACrefYearMonthDay{2017}{}{}.
\newblock
{\BBOQ}\APACrefatitle {Potential links between continental rifting, CO2
  degassing and climate change through time} {Potential links between
  continental rifting, co2 degassing and climate change through time}.{\BBCQ}
\newblock
\APACjournalVolNumPages{Nature Geoscience}{10}{12}{941--946}.
\PrintBackRefs{\CurrentBib}

\bibitem [\protect \citeauthoryear {%
Cox%
\ \protect \BOthers {.}}{%
Cox%
\ \protect \BOthers {.}}{%
{\protect \APACyear {2016}}%
}]{%
Cox-Halverson-Stevenson-et-al-2016:continental}
\APACinsertmetastar {%
Cox-Halverson-Stevenson-et-al-2016:continental}%
\begin{APACrefauthors}%
Cox, G\BPBI M.%
, Halverson, G\BPBI P.%
, Stevenson, R\BPBI K.%
, Vokaty, M.%
, Poirier, A.%
, Kunzmann, M.%
\BDBL {}Macdonald, F\BPBI A.%
\end{APACrefauthors}%
\unskip\
\newblock
\APACrefYearMonthDay{2016}{}{}.
\newblock
{\BBOQ}\APACrefatitle {Continental flood basalt weathering as a trigger for
  {Neoproterozoic} {Snowball Earth}} {Continental flood basalt weathering as a
  trigger for {Neoproterozoic} {Snowball Earth}}.{\BBCQ}
\newblock
\APACjournalVolNumPages{Earth and Planetary Science Letters}{446}{}{89--99}.
\PrintBackRefs{\CurrentBib}

\bibitem [\protect \citeauthoryear {%
Donnadieu%
\ \protect \BOthers {.}}{%
Donnadieu%
\ \protect \BOthers {.}}{%
{\protect \APACyear {2006}}%
}]{%
Donnadieu-Godderis-Pierrehumbert-et-al-2006:geoclim}
\APACinsertmetastar {%
Donnadieu-Godderis-Pierrehumbert-et-al-2006:geoclim}%
\begin{APACrefauthors}%
Donnadieu, Y.%
, Godd{\'e}ris, Y.%
, Pierrehumbert, R.%
, Dromart, G.%
, Fluteau, F.%
\BCBL {}\ \BBA {} Jacob, R.%
\end{APACrefauthors}%
\unskip\
\newblock
\APACrefYearMonthDay{2006}{}{}.
\newblock
{\BBOQ}\APACrefatitle {A {GEOCLIM} simulation of climatic and biogeochemical
  consequences of {Pangea} breakup} {A {GEOCLIM} simulation of climatic and
  biogeochemical consequences of {Pangea} breakup}.{\BBCQ}
\newblock
\APACjournalVolNumPages{Geochemistry, Geophysics, Geosystems}{7}{11}{}.
\PrintBackRefs{\CurrentBib}

\bibitem [\protect \citeauthoryear {%
Donnadieu%
, Godd{\'e}ris%
, Ramstein%
, N{\'e}d{\'e}lec%
\BCBL {}\ \BBA {} Meert%
}{%
Donnadieu%
, Godd{\'e}ris%
\BCBL {}\ \protect \BOthers {.}}{%
{\protect \APACyear {2004}}%
}]{%
Donnadieu-Godderis-Ramstein-et-al-2004:snowball}
\APACinsertmetastar {%
Donnadieu-Godderis-Ramstein-et-al-2004:snowball}%
\begin{APACrefauthors}%
Donnadieu, Y.%
, Godd{\'e}ris, Y.%
, Ramstein, G.%
, N{\'e}d{\'e}lec, A.%
\BCBL {}\ \BBA {} Meert, J.%
\end{APACrefauthors}%
\unskip\
\newblock
\APACrefYearMonthDay{2004}{}{}.
\newblock
{\BBOQ}\APACrefatitle {A `snowball {Earth}’ climate triggered by continental
  break-up through changes in runoff} {A `snowball {Earth}’ climate triggered
  by continental break-up through changes in runoff}.{\BBCQ}
\newblock
\APACjournalVolNumPages{Nature}{428}{6980}{303--306}.
\PrintBackRefs{\CurrentBib}

\bibitem [\protect \citeauthoryear {%
Donnadieu%
, Ramstein%
, Fluteau%
, Roche%
\BCBL {}\ \BBA {} Ganopolski%
}{%
Donnadieu%
, Ramstein%
\BCBL {}\ \protect \BOthers {.}}{%
{\protect \APACyear {2004}}%
}]{%
Donnadieu-Ramstein-Fluteau-et-al-2004:impact}
\APACinsertmetastar {%
Donnadieu-Ramstein-Fluteau-et-al-2004:impact}%
\begin{APACrefauthors}%
Donnadieu, Y.%
, Ramstein, G.%
, Fluteau, F.%
, Roche, D.%
\BCBL {}\ \BBA {} Ganopolski, A.%
\end{APACrefauthors}%
\unskip\
\newblock
\APACrefYearMonthDay{2004}{}{}.
\newblock
{\BBOQ}\APACrefatitle {The impact of atmospheric and oceanic heat transports on
  the sea-ice-albedo instability during the {Neoproterozoic}} {The impact of
  atmospheric and oceanic heat transports on the sea-ice-albedo instability
  during the {Neoproterozoic}}.{\BBCQ}
\newblock
\APACjournalVolNumPages{Climate Dynamics}{22}{2}{293--306}.
\PrintBackRefs{\CurrentBib}

\bibitem [\protect \citeauthoryear {%
Edmond%
\ \BBA {} Huh%
}{%
Edmond%
\ \BBA {} Huh%
}{%
{\protect \APACyear {2003}}%
}]{%
Edmond-Huh-2003:non}
\APACinsertmetastar {%
Edmond-Huh-2003:non}%
\begin{APACrefauthors}%
Edmond, J\BPBI M.%
\BCBT {}\ \BBA {} Huh, Y.%
\end{APACrefauthors}%
\unskip\
\newblock
\APACrefYearMonthDay{2003}{}{}.
\newblock
{\BBOQ}\APACrefatitle {Non-steady state carbonate recycling and implications
  for the evolution of atmospheric {P}$_{{\mathrm{CO}}_2}$} {Non-steady state
  carbonate recycling and implications for the evolution of atmospheric
  {P}$_{{\mathrm{CO}}_2}$}.{\BBCQ}
\newblock
\APACjournalVolNumPages{Earth and Planetary Science
  Letters}{216}{1-2}{125--139}.
\PrintBackRefs{\CurrentBib}

\bibitem [\protect \citeauthoryear {%
Evans%
}{%
Evans%
}{%
{\protect \APACyear {2013}}%
}]{%
Evans-2013:reconstructing}
\APACinsertmetastar {%
Evans-2013:reconstructing}%
\begin{APACrefauthors}%
Evans, D\BPBI A.%
\end{APACrefauthors}%
\unskip\
\newblock
\APACrefYearMonthDay{2013}{11}{}.
\newblock
{\BBOQ}\APACrefatitle {Reconstructing {pre-Pangean} supercontinents}
  {Reconstructing {pre-Pangean} supercontinents}.{\BBCQ}
\newblock
\APACjournalVolNumPages{GSA Bulletin}{125}{11-12}{1735-1751}.
\newblock
\begin{APACrefDOI} \doi{10.1130/B30950.1} \end{APACrefDOI}
\PrintBackRefs{\CurrentBib}

\bibitem [\protect \citeauthoryear {%
Evans%
, Beukes%
\BCBL {}\ \BBA {} Kirschvink%
}{%
Evans%
\ \protect \BOthers {.}}{%
{\protect \APACyear {1997}}%
}]{%
Evans-Beukes-Kirschvink-1997:low}
\APACinsertmetastar {%
Evans-Beukes-Kirschvink-1997:low}%
\begin{APACrefauthors}%
Evans, D\BPBI A.%
, Beukes, N\BPBI J.%
\BCBL {}\ \BBA {} Kirschvink, J\BPBI L.%
\end{APACrefauthors}%
\unskip\
\newblock
\APACrefYearMonthDay{1997}{}{}.
\newblock
{\BBOQ}\APACrefatitle {Low-latitude glaciation in the {Palaeoproterozoic} era}
  {Low-latitude glaciation in the {Palaeoproterozoic} era}.{\BBCQ}
\newblock
\APACjournalVolNumPages{Nature}{386}{6622}{262--266}.
\PrintBackRefs{\CurrentBib}

\bibitem [\protect \citeauthoryear {%
Gerlach%
}{%
Gerlach%
}{%
{\protect \APACyear {2011}}%
}]{%
Gerlach-2011:volcanic}
\APACinsertmetastar {%
Gerlach-2011:volcanic}%
\begin{APACrefauthors}%
Gerlach, T.%
\end{APACrefauthors}%
\unskip\
\newblock
\APACrefYearMonthDay{2011}{}{}.
\newblock
{\BBOQ}\APACrefatitle {Volcanic versus anthropogenic carbon dioxide} {Volcanic
  versus anthropogenic carbon dioxide}.{\BBCQ}
\newblock
\APACjournalVolNumPages{Eos, Transactions American Geophysical
  Union}{92}{24}{201--202}.
\PrintBackRefs{\CurrentBib}

\bibitem [\protect \citeauthoryear {%
Godd{\'e}ris%
\ \BBA {} Donnadieu%
}{%
Godd{\'e}ris%
\ \BBA {} Donnadieu%
}{%
{\protect \APACyear {2019}}%
}]{%
Godderis-Donnadieu-2019:sink}
\APACinsertmetastar {%
Godderis-Donnadieu-2019:sink}%
\begin{APACrefauthors}%
Godd{\'e}ris, Y.%
\BCBT {}\ \BBA {} Donnadieu, Y.%
\end{APACrefauthors}%
\unskip\
\newblock
\APACrefYearMonthDay{2019}{}{}.
\newblock
{\BBOQ}\APACrefatitle {A sink-or a source-driven carbon cycle at the geological
  timescale? {Relative} importance of palaeogeography versus solid {Earth}
  degassing rate in the {Phanerozoic} climatic evolution} {A sink-or a
  source-driven carbon cycle at the geological timescale? {Relative} importance
  of palaeogeography versus solid {Earth} degassing rate in the {Phanerozoic}
  climatic evolution}.{\BBCQ}
\newblock
\APACjournalVolNumPages{Geological Magazine}{156}{2}{355--365}.
\PrintBackRefs{\CurrentBib}

\bibitem [\protect \citeauthoryear {%
Godd{\'e}ris%
\ \protect \BOthers {.}}{%
Godd{\'e}ris%
\ \protect \BOthers {.}}{%
{\protect \APACyear {2017}}%
}]{%
Godderis-Donnadieu-Carretier-et-al-2017:onset}
\APACinsertmetastar {%
Godderis-Donnadieu-Carretier-et-al-2017:onset}%
\begin{APACrefauthors}%
Godd{\'e}ris, Y.%
, Donnadieu, Y.%
, Carretier, S.%
, Aretz, M.%
, Dera, G.%
, Macouin, M.%
\BCBL {}\ \BBA {} Regard, V.%
\end{APACrefauthors}%
\unskip\
\newblock
\APACrefYearMonthDay{2017}{}{}.
\newblock
{\BBOQ}\APACrefatitle {Onset and ending of the late {Palaeozoic} ice age
  triggered by tectonically paced rock weathering} {Onset and ending of the
  late {Palaeozoic} ice age triggered by tectonically paced rock
  weathering}.{\BBCQ}
\newblock
\APACjournalVolNumPages{Nature Geoscience}{10}{5}{382--386}.
\PrintBackRefs{\CurrentBib}

\bibitem [\protect \citeauthoryear {%
Godd{\'e}ris%
\ \protect \BOthers {.}}{%
Godd{\'e}ris%
\ \protect \BOthers {.}}{%
{\protect \APACyear {2003}}%
}]{%
Godderis-Donnadieu-Nedelec-et-al-2003:sturtian}
\APACinsertmetastar {%
Godderis-Donnadieu-Nedelec-et-al-2003:sturtian}%
\begin{APACrefauthors}%
Godd{\'e}ris, Y.%
, Donnadieu, Y.%
, N{\'e}d{\'e}lec, A.%
, Dupr{\'e}, B.%
, Dessert, C.%
, Grard, A.%
\BDBL {}Fran{\c{c}}ois, L.%
\end{APACrefauthors}%
\unskip\
\newblock
\APACrefYearMonthDay{2003}{}{}.
\newblock
{\BBOQ}\APACrefatitle {The {Sturtian} `snowball' glaciation: {Fire} and ice}
  {The {Sturtian} `snowball' glaciation: {Fire} and ice}.{\BBCQ}
\newblock
\APACjournalVolNumPages{Earth and Planetary Science Letters}{211}{1-2}{1--12}.
\PrintBackRefs{\CurrentBib}

\bibitem [\protect \citeauthoryear {%
Graham%
\ \BBA {} Pierrehumbert%
}{%
Graham%
\ \BBA {} Pierrehumbert%
}{%
{\protect \APACyear {2020}}%
}]{%
Graham-Pierrehumbert-2020:thermodynamic}
\APACinsertmetastar {%
Graham-Pierrehumbert-2020:thermodynamic}%
\begin{APACrefauthors}%
Graham, R\BPBI J.%
\BCBT {}\ \BBA {} Pierrehumbert, R.%
\end{APACrefauthors}%
\unskip\
\newblock
\APACrefYearMonthDay{2020}{}{}.
\newblock
{\BBOQ}\APACrefatitle {Thermodynamic and energetic limits on continental
  silicate weathering strongly impact the climate and habitability of wet,
  rocky worlds} {Thermodynamic and energetic limits on continental silicate
  weathering strongly impact the climate and habitability of wet, rocky
  worlds}.{\BBCQ}
\newblock
\APACjournalVolNumPages{The Astrophysical Journal}{896}{2}{115}.
\PrintBackRefs{\CurrentBib}

\bibitem [\protect \citeauthoryear {%
Greenwood%
\ \BBA {} Wing%
}{%
Greenwood%
\ \BBA {} Wing%
}{%
{\protect \APACyear {1995}}%
}]{%
Greenwood-Wing-1995:eocene}
\APACinsertmetastar {%
Greenwood-Wing-1995:eocene}%
\begin{APACrefauthors}%
Greenwood, D\BPBI R.%
\BCBT {}\ \BBA {} Wing, S\BPBI L.%
\end{APACrefauthors}%
\unskip\
\newblock
\APACrefYearMonthDay{1995}{}{}.
\newblock
{\BBOQ}\APACrefatitle {{Eocene} continental climates and latitudinal
  temperature gradients} {{Eocene} continental climates and latitudinal
  temperature gradients}.{\BBCQ}
\newblock
\APACjournalVolNumPages{Geology}{23}{11}{1044--1048}.
\PrintBackRefs{\CurrentBib}

\bibitem [\protect \citeauthoryear {%
Haqq-Misra%
, Kopparapu%
, Batalha%
, Harman%
\BCBL {}\ \BBA {} Kasting%
}{%
Haqq-Misra%
\ \protect \BOthers {.}}{%
{\protect \APACyear {2016}}%
}]{%
Haqq-Misra-Kopparapu-Batalha-et-al-2016:limit}
\APACinsertmetastar {%
Haqq-Misra-Kopparapu-Batalha-et-al-2016:limit}%
\begin{APACrefauthors}%
Haqq-Misra, J.%
, Kopparapu, R\BPBI K.%
, Batalha, N\BPBI E.%
, Harman, C\BPBI E.%
\BCBL {}\ \BBA {} Kasting, J\BPBI F.%
\end{APACrefauthors}%
\unskip\
\newblock
\APACrefYearMonthDay{2016}{}{}.
\newblock
{\BBOQ}\APACrefatitle {Limit cycles can reduce the width of the habitable zone}
  {Limit cycles can reduce the width of the habitable zone}.{\BBCQ}
\newblock
\APACjournalVolNumPages{The Astrophysical Journal}{827}{2}{120}.
\PrintBackRefs{\CurrentBib}

\bibitem [\protect \citeauthoryear {%
Hoffman%
, Kaufman%
, Halverson%
\BCBL {}\ \BBA {} Schrag%
}{%
Hoffman%
\ \protect \BOthers {.}}{%
{\protect \APACyear {1998}}%
}]{%
Hoffman-Kaufman-Halverson-et-al-1998:neoproterozoic}
\APACinsertmetastar {%
Hoffman-Kaufman-Halverson-et-al-1998:neoproterozoic}%
\begin{APACrefauthors}%
Hoffman, P\BPBI F.%
, Kaufman, A\BPBI J.%
, Halverson, G\BPBI P.%
\BCBL {}\ \BBA {} Schrag, D\BPBI P.%
\end{APACrefauthors}%
\unskip\
\newblock
\APACrefYearMonthDay{1998}{aug 28}{}.
\newblock
{\BBOQ}\APACrefatitle {A {Neoproterozoic} snowball {Earth}} {A {Neoproterozoic}
  snowball {Earth}}.{\BBCQ}
\newblock
\APACjournalVolNumPages{Science}{281}{5381}{1342-1346}.
\PrintBackRefs{\CurrentBib}

\bibitem [\protect \citeauthoryear {%
Hoffman%
\ \BBA {} Schrag%
}{%
Hoffman%
\ \BBA {} Schrag%
}{%
{\protect \APACyear {2002}}%
}]{%
Hoffman-Schrag-2002:snowball}
\APACinsertmetastar {%
Hoffman-Schrag-2002:snowball}%
\begin{APACrefauthors}%
Hoffman, P\BPBI F.%
\BCBT {}\ \BBA {} Schrag, D\BPBI P.%
\end{APACrefauthors}%
\unskip\
\newblock
\APACrefYearMonthDay{2002}{JUN}{}.
\newblock
{\BBOQ}\APACrefatitle {The snowball {Earth} hypothesis: testing the limits of
  global change} {The snowball {Earth} hypothesis: testing the limits of global
  change}.{\BBCQ}
\newblock
\APACjournalVolNumPages{Terra Nova}{14}{3}{129-155}.
\newblock
\begin{APACrefDOI} \doi{10.1046/j.1365-3121.2002.00408.x} \end{APACrefDOI}
\PrintBackRefs{\CurrentBib}

\bibitem [\protect \citeauthoryear {%
Hunter%
}{%
Hunter%
}{%
{\protect \APACyear {2007}}%
}]{%
Hunter:2007}
\APACinsertmetastar {%
Hunter:2007}%
\begin{APACrefauthors}%
Hunter, J\BPBI D.%
\end{APACrefauthors}%
\unskip\
\newblock
\APACrefYearMonthDay{2007}{}{}.
\newblock
{\BBOQ}\APACrefatitle {Matplotlib: A 2D graphics environment} {Matplotlib: A 2d
  graphics environment}.{\BBCQ}
\newblock
\APACjournalVolNumPages{Computing in Science \& Engineering}{9}{3}{90--95}.
\newblock
\begin{APACrefDOI} \doi{10.1109/MCSE.2007.55} \end{APACrefDOI}
\PrintBackRefs{\CurrentBib}

\bibitem [\protect \citeauthoryear {%
Kirschvink%
\ \protect \BOthers {.}}{%
Kirschvink%
\ \protect \BOthers {.}}{%
{\protect \APACyear {2000}}%
}]{%
Kirschvink-Gaidos-Bertani-et-al-2000:paleoproterozoic}
\APACinsertmetastar {%
Kirschvink-Gaidos-Bertani-et-al-2000:paleoproterozoic}%
\begin{APACrefauthors}%
Kirschvink, J\BPBI L.%
, Gaidos, E\BPBI J.%
, Bertani, L\BPBI E.%
, Beukes, N\BPBI J.%
, Gutzmer, J.%
, Maepa, L\BPBI N.%
\BCBL {}\ \BBA {} Steinberger, R\BPBI E.%
\end{APACrefauthors}%
\unskip\
\newblock
\APACrefYearMonthDay{2000}{}{}.
\newblock
{\BBOQ}\APACrefatitle {Paleoproterozoic snowball Earth: {Extreme} climatic and
  geochemical global change and its biological consequences} {Paleoproterozoic
  snowball earth: {Extreme} climatic and geochemical global change and its
  biological consequences}.{\BBCQ}
\newblock
\APACjournalVolNumPages{Proceedings of the National Academy of
  Sciences}{97}{4}{1400--1405}.
\PrintBackRefs{\CurrentBib}

\bibitem [\protect \citeauthoryear {%
Lagache%
}{%
Lagache%
}{%
{\protect \APACyear {1976}}%
}]{%
Lagache-1976:new}
\APACinsertmetastar {%
Lagache-1976:new}%
\begin{APACrefauthors}%
Lagache, M.%
\end{APACrefauthors}%
\unskip\
\newblock
\APACrefYearMonthDay{1976}{}{}.
\newblock
{\BBOQ}\APACrefatitle {New data on the kinetics of the dissolution of alkali
  feldspars at 200 {\textdegree}{C} in {CO}$_2$ charged water} {New data on the
  kinetics of the dissolution of alkali feldspars at 200 {\textdegree}{C} in
  {CO}$_2$ charged water}.{\BBCQ}
\newblock
\APACjournalVolNumPages{Geochimica et Cosmochimica Acta}{40}{2}{157--161}.
\PrintBackRefs{\CurrentBib}

\bibitem [\protect \citeauthoryear {%
Lawrence%
\ \protect \BOthers {.}}{%
Lawrence%
\ \protect \BOthers {.}}{%
{\protect \APACyear {2011}}%
}]{%
Lawrence-Oleson-Flanner-et-al-2011:parameterization}
\APACinsertmetastar {%
Lawrence-Oleson-Flanner-et-al-2011:parameterization}%
\begin{APACrefauthors}%
Lawrence, D\BPBI M.%
, Oleson, K\BPBI W.%
, Flanner, M\BPBI G.%
, Thornton, P\BPBI E.%
, Swenson, S\BPBI C.%
, Lawrence, P\BPBI J.%
\BDBL {}Slater, A\BPBI G.%
\end{APACrefauthors}%
\unskip\
\newblock
\APACrefYearMonthDay{2011}{}{}.
\newblock
{\BBOQ}\APACrefatitle {Parameterization improvements and functional and
  structural advances in {Version} 4 of the {Community Land Model}}
  {Parameterization improvements and functional and structural advances in
  {Version} 4 of the {Community Land Model}}.{\BBCQ}
\newblock
\APACjournalVolNumPages{Journal of Advances in Modeling Earth Systems}{3}{1}{}.
\PrintBackRefs{\CurrentBib}

\bibitem [\protect \citeauthoryear {%
Li%
\ \protect \BOthers {.}}{%
Li%
\ \protect \BOthers {.}}{%
{\protect \APACyear {2008}}%
}]{%
Li-Bogdanova-Collins-et-al-2008:assembly}
\APACinsertmetastar {%
Li-Bogdanova-Collins-et-al-2008:assembly}%
\begin{APACrefauthors}%
Li, Z\BPBI X.%
, Bogdanova, S\BPBI V.%
, Collins, A\BPBI S.%
, Davidson, A.%
, De~Waele, B.%
, Ernst, R\BPBI E.%
\BDBL {}Vernikovsky, V.%
\end{APACrefauthors}%
\unskip\
\newblock
\APACrefYearMonthDay{2008}{}{}.
\newblock
{\BBOQ}\APACrefatitle {Assembly, configuration, and break-up history of
  {Rodinia}: {A} synthesis} {Assembly, configuration, and break-up history of
  {Rodinia}: {A} synthesis}.{\BBCQ}
\newblock
\APACjournalVolNumPages{Precambrian Research}{160}{}{179-210}.
\PrintBackRefs{\CurrentBib}

\bibitem [\protect \citeauthoryear {%
Macdonald%
\ \protect \BOthers {.}}{%
Macdonald%
\ \protect \BOthers {.}}{%
{\protect \APACyear {2010}}%
}]{%
Macdonald-Schmitz-Crowley-et-al-2010:calibrating}
\APACinsertmetastar {%
Macdonald-Schmitz-Crowley-et-al-2010:calibrating}%
\begin{APACrefauthors}%
Macdonald, F\BPBI A.%
, Schmitz, M\BPBI D.%
, Crowley, J\BPBI L.%
, Roots, C\BPBI F.%
, Jones, D\BPBI S.%
, Maloof, A\BPBI C.%
\BDBL {}Schrag, D\BPBI P.%
\end{APACrefauthors}%
\unskip\
\newblock
\APACrefYearMonthDay{2010}{MAR 5}{}.
\newblock
{\BBOQ}\APACrefatitle {Calibrating the {Cryogenian}} {Calibrating the
  {Cryogenian}}.{\BBCQ}
\newblock
\APACjournalVolNumPages{Science}{327}{5970}{1241-1243}.
\newblock
\begin{APACrefDOI} \doi{10.1126/science.1183325} \end{APACrefDOI}
\PrintBackRefs{\CurrentBib}

\bibitem [\protect \citeauthoryear {%
Macdonald%
, Swanson-Hysell%
, Park%
, Lisiecki%
\BCBL {}\ \BBA {} Jagoutz%
}{%
Macdonald%
\ \protect \BOthers {.}}{%
{\protect \APACyear {2019}}%
}]{%
Macdonald-Swanson-Hysell-Park-et-al-2019:arc}
\APACinsertmetastar {%
Macdonald-Swanson-Hysell-Park-et-al-2019:arc}%
\begin{APACrefauthors}%
Macdonald, F\BPBI A.%
, Swanson-Hysell, N\BPBI L.%
, Park, Y.%
, Lisiecki, L.%
\BCBL {}\ \BBA {} Jagoutz, O.%
\end{APACrefauthors}%
\unskip\
\newblock
\APACrefYearMonthDay{2019}{}{}.
\newblock
{\BBOQ}\APACrefatitle {Arc-continent collisions in the tropics set {Earth's}
  climate state} {Arc-continent collisions in the tropics set {Earth's} climate
  state}.{\BBCQ}
\newblock
\APACjournalVolNumPages{Science}{364}{6436}{181--184}.
\PrintBackRefs{\CurrentBib}

\bibitem [\protect \citeauthoryear {%
Maher%
\ \BBA {} Chamberlain%
}{%
Maher%
\ \BBA {} Chamberlain%
}{%
{\protect \APACyear {2014}}%
}]{%
Maher-Chamberlain-2014:hydrologic}
\APACinsertmetastar {%
Maher-Chamberlain-2014:hydrologic}%
\begin{APACrefauthors}%
Maher, K.%
\BCBT {}\ \BBA {} Chamberlain, C\BPBI P.%
\end{APACrefauthors}%
\unskip\
\newblock
\APACrefYearMonthDay{2014}{}{}.
\newblock
{\BBOQ}\APACrefatitle {Hydrologic regulation of chemical weathering and the
  geologic carbon cycle} {Hydrologic regulation of chemical weathering and the
  geologic carbon cycle}.{\BBCQ}
\newblock
\APACjournalVolNumPages{Science}{343}{6178}{1502--1504}.
\PrintBackRefs{\CurrentBib}

\bibitem [\protect \citeauthoryear {%
Maher%
, Steefel%
, DePaolo%
\BCBL {}\ \BBA {} Viani%
}{%
Maher%
\ \protect \BOthers {.}}{%
{\protect \APACyear {2006}}%
}]{%
Maher-Steefel-DePaolo-et-al-2006:mineral}
\APACinsertmetastar {%
Maher-Steefel-DePaolo-et-al-2006:mineral}%
\begin{APACrefauthors}%
Maher, K.%
, Steefel, C\BPBI I.%
, DePaolo, D\BPBI J.%
\BCBL {}\ \BBA {} Viani, B\BPBI E.%
\end{APACrefauthors}%
\unskip\
\newblock
\APACrefYearMonthDay{2006}{}{}.
\newblock
{\BBOQ}\APACrefatitle {The mineral dissolution rate conundrum: {Insights} from
  reactive transport modeling of {U} isotopes and pore fluid chemistry in
  marine sediments} {The mineral dissolution rate conundrum: {Insights} from
  reactive transport modeling of {U} isotopes and pore fluid chemistry in
  marine sediments}.{\BBCQ}
\newblock
\APACjournalVolNumPages{Geochimica et Cosmochimica Acta}{70}{2}{337--363}.
\PrintBackRefs{\CurrentBib}

\bibitem [\protect \citeauthoryear {%
Maher%
, Steefel%
, White%
\BCBL {}\ \BBA {} Stonestrom%
}{%
Maher%
\ \protect \BOthers {.}}{%
{\protect \APACyear {2009}}%
}]{%
Maher-Steefel-White-et-al-2009:role}
\APACinsertmetastar {%
Maher-Steefel-White-et-al-2009:role}%
\begin{APACrefauthors}%
Maher, K.%
, Steefel, C\BPBI I.%
, White, A\BPBI F.%
\BCBL {}\ \BBA {} Stonestrom, D\BPBI A.%
\end{APACrefauthors}%
\unskip\
\newblock
\APACrefYearMonthDay{2009}{}{}.
\newblock
{\BBOQ}\APACrefatitle {The role of reaction affinity and secondary minerals in
  regulating chemical weathering rates at the {Santa Cruz Soil Chronosequence},
  {California}} {The role of reaction affinity and secondary minerals in
  regulating chemical weathering rates at the {Santa Cruz Soil Chronosequence},
  {California}}.{\BBCQ}
\newblock
\APACjournalVolNumPages{Geochimica et Cosmochimica Acta}{73}{10}{2804--2831}.
\PrintBackRefs{\CurrentBib}

\bibitem [\protect \citeauthoryear {%
Manabe%
\ \BBA {} Wetherald%
}{%
Manabe%
\ \BBA {} Wetherald%
}{%
{\protect \APACyear {1975}}%
}]{%
Manabe-Wetherald-1975:effects}
\APACinsertmetastar {%
Manabe-Wetherald-1975:effects}%
\begin{APACrefauthors}%
Manabe, S.%
\BCBT {}\ \BBA {} Wetherald, R\BPBI T.%
\end{APACrefauthors}%
\unskip\
\newblock
\APACrefYearMonthDay{1975}{}{}.
\newblock
{\BBOQ}\APACrefatitle {The effects of doubling the {CO}$_2$ concentration on
  the climate of a general circulation model} {The effects of doubling the
  {CO}$_2$ concentration on the climate of a general circulation model}.{\BBCQ}
\newblock
\APACjournalVolNumPages{Journal of Atmospheric Sciences}{32}{1}{3--15}.
\PrintBackRefs{\CurrentBib}

\bibitem [\protect \citeauthoryear {%
Marshall%
, Walker%
\BCBL {}\ \BBA {} Kuhn%
}{%
Marshall%
\ \protect \BOthers {.}}{%
{\protect \APACyear {1988}}%
}]{%
Marshall-Walker-Kuhn-1988:long}
\APACinsertmetastar {%
Marshall-Walker-Kuhn-1988:long}%
\begin{APACrefauthors}%
Marshall, H\BPBI G.%
, Walker, J\BPBI C\BPBI G.%
\BCBL {}\ \BBA {} Kuhn, W\BPBI R.%
\end{APACrefauthors}%
\unskip\
\newblock
\APACrefYearMonthDay{1988}{}{}.
\newblock
{\BBOQ}\APACrefatitle {Long-term climate change and the geochemical cycle of
  carbon} {Long-term climate change and the geochemical cycle of
  carbon}.{\BBCQ}
\newblock
\APACjournalVolNumPages{Journal of Geophysical Research:
  Atmospheres}{93}{D1}{791--801}.
\PrintBackRefs{\CurrentBib}

\bibitem [\protect \citeauthoryear {%
Neale%
\ \protect \BOthers {.}}{%
Neale%
\ \protect \BOthers {.}}{%
{\protect \APACyear {2012}}%
}]{%
Neale-Chen-Gettelman-et-al-2012:description}
\APACinsertmetastar {%
Neale-Chen-Gettelman-et-al-2012:description}%
\begin{APACrefauthors}%
Neale, R\BPBI B.%
, Chen, C\BHBI C.%
, Gettelman, A.%
, Lauritzen, P\BPBI H.%
, Park, S.%
, Williamson, D\BPBI L.%
\BDBL {}others%
\end{APACrefauthors}%
\unskip\
\newblock
\APACrefYearMonthDay{2012}{}{}.
\newblock
{\BBOQ}\APACrefatitle {Description of the {NCAR} community atmosphere model
  ({CAM 5.0})} {Description of the {NCAR} community atmosphere model ({CAM
  5.0})}.{\BBCQ}
\newblock
\APACjournalVolNumPages{NCAR Tech. Note NCAR/TN-486+ STR}{}{}{}.
\PrintBackRefs{\CurrentBib}

\bibitem [\protect \citeauthoryear {%
Park%
\ \protect \BOthers {.}}{%
Park%
\ \protect \BOthers {.}}{%
{\protect \APACyear {2020}}%
}]{%
park2020emergence}
\APACinsertmetastar {%
park2020emergence}%
\begin{APACrefauthors}%
Park, Y.%
, Maffre, P.%
, Godd{\'e}ris, Y.%
, Macdonald, F\BPBI A.%
, Anttila, E\BPBI S.%
\BCBL {}\ \BBA {} Swanson-Hysell, N\BPBI L.%
\end{APACrefauthors}%
\unskip\
\newblock
\APACrefYearMonthDay{2020}{}{}.
\newblock
{\BBOQ}\APACrefatitle {Emergence of the Southeast Asian islands as a driver for
  Neogene cooling} {Emergence of the southeast asian islands as a driver for
  neogene cooling}.{\BBCQ}
\newblock
\APACjournalVolNumPages{Proceedings of the National Academy of
  Sciences}{117}{41}{25319--25326}.
\PrintBackRefs{\CurrentBib}

\bibitem [\protect \citeauthoryear {%
Pierrehumbert%
, Abbot%
, Voigt%
\BCBL {}\ \BBA {} Koll%
}{%
Pierrehumbert%
\ \protect \BOthers {.}}{%
{\protect \APACyear {2011}}%
}]{%
Pierrehumbert-Abbot-Voigt-et-al-2011:climate}
\APACinsertmetastar {%
Pierrehumbert-Abbot-Voigt-et-al-2011:climate}%
\begin{APACrefauthors}%
Pierrehumbert, R.%
, Abbot, D.%
, Voigt, A.%
\BCBL {}\ \BBA {} Koll, D.%
\end{APACrefauthors}%
\unskip\
\newblock
\APACrefYearMonthDay{2011}{}{}.
\newblock
{\BBOQ}\APACrefatitle {Climate of the {Neoproterozoic}} {Climate of the
  {Neoproterozoic}}.{\BBCQ}
\newblock
\APACjournalVolNumPages{Annual Review of Earth and Planetary
  Sciences}{39}{}{417--460}.
\PrintBackRefs{\CurrentBib}

\bibitem [\protect \citeauthoryear {%
Prave%
, Condon%
, Hoffmann%
, Tapster%
\BCBL {}\ \BBA {} Fallick%
}{%
Prave%
\ \protect \BOthers {.}}{%
{\protect \APACyear {2016}}%
}]{%
Prave-Condon-Hoffmann-et-al-2016:duration}
\APACinsertmetastar {%
Prave-Condon-Hoffmann-et-al-2016:duration}%
\begin{APACrefauthors}%
Prave, A\BPBI R.%
, Condon, D\BPBI J.%
, Hoffmann, K\BPBI H.%
, Tapster, S.%
\BCBL {}\ \BBA {} Fallick, A\BPBI E.%
\end{APACrefauthors}%
\unskip\
\newblock
\APACrefYearMonthDay{2016}{}{}.
\newblock
{\BBOQ}\APACrefatitle {Duration and nature of the {end-Cryogenian} ({Marinoan})
  glaciation} {Duration and nature of the {end-Cryogenian} ({Marinoan})
  glaciation}.{\BBCQ}
\newblock
\APACjournalVolNumPages{Geology}{44}{8}{631--634}.
\PrintBackRefs{\CurrentBib}

\bibitem [\protect \citeauthoryear {%
Rooney%
\ \protect \BOthers {.}}{%
Rooney%
\ \protect \BOthers {.}}{%
{\protect \APACyear {2014}}%
}]{%
Rooney-Macdonald-Strauss-et-al-2014:re}
\APACinsertmetastar {%
Rooney-Macdonald-Strauss-et-al-2014:re}%
\begin{APACrefauthors}%
Rooney, A\BPBI D.%
, Macdonald, F\BPBI A.%
, Strauss, J\BPBI V.%
, Dud{\'a}s, F\BPBI {\"O}.%
, Hallmann, C.%
\BCBL {}\ \BBA {} Selby, D.%
\end{APACrefauthors}%
\unskip\
\newblock
\APACrefYearMonthDay{2014}{}{}.
\newblock
{\BBOQ}\APACrefatitle {{Re-Os} geochronology and coupled Os-Sr isotope
  constraints on the {Sturtian} snowball {Earth}} {{Re-Os} geochronology and
  coupled os-sr isotope constraints on the {Sturtian} snowball {Earth}}.{\BBCQ}
\newblock
\APACjournalVolNumPages{Proceedings of the National Academy of
  Sciences}{111}{1}{51--56}.
\PrintBackRefs{\CurrentBib}

\bibitem [\protect \citeauthoryear {%
Rooney%
, Strauss%
, Brandon%
\BCBL {}\ \BBA {} Macdonald%
}{%
Rooney%
\ \protect \BOthers {.}}{%
{\protect \APACyear {2015}}%
}]{%
Rooney-Strauss-Brandon-et-al-2015:cryogenian}
\APACinsertmetastar {%
Rooney-Strauss-Brandon-et-al-2015:cryogenian}%
\begin{APACrefauthors}%
Rooney, A\BPBI D.%
, Strauss, J\BPBI V.%
, Brandon, A\BPBI D.%
\BCBL {}\ \BBA {} Macdonald, F\BPBI A.%
\end{APACrefauthors}%
\unskip\
\newblock
\APACrefYearMonthDay{2015}{}{}.
\newblock
{\BBOQ}\APACrefatitle {A {Cryogenian} chronology: {Two} long-lasting
  synchronous {Neoproterozoic} glaciations} {A {Cryogenian} chronology: {Two}
  long-lasting synchronous {Neoproterozoic} glaciations}.{\BBCQ}
\newblock
\APACjournalVolNumPages{Geology}{43}{5}{459--462}.
\PrintBackRefs{\CurrentBib}

\bibitem [\protect \citeauthoryear {%
Schrag%
, Berner%
, Hoffman%
\BCBL {}\ \BBA {} Halverson%
}{%
Schrag%
\ \protect \BOthers {.}}{%
{\protect \APACyear {2002}}%
}]{%
Schrag-Berner-Hoffman-et-al-2002:initiation}
\APACinsertmetastar {%
Schrag-Berner-Hoffman-et-al-2002:initiation}%
\begin{APACrefauthors}%
Schrag, D\BPBI P.%
, Berner, R\BPBI A.%
, Hoffman, P\BPBI F.%
\BCBL {}\ \BBA {} Halverson, G\BPBI P.%
\end{APACrefauthors}%
\unskip\
\newblock
\APACrefYearMonthDay{2002}{JUN 27}{}.
\newblock
{\BBOQ}\APACrefatitle {On the initiation of a snowball {Earth}} {On the
  initiation of a snowball {Earth}}.{\BBCQ}
\newblock
\APACjournalVolNumPages{Geochemistry Geophysics Geosystems}{3}{}{}.
\newblock
\begin{APACrefDOI} \doi{10.1029/2001GC000219} \end{APACrefDOI}
\PrintBackRefs{\CurrentBib}

\bibitem [\protect \citeauthoryear {%
Seabold%
\ \BBA {} Perktold%
}{%
Seabold%
\ \BBA {} Perktold%
}{%
{\protect \APACyear {2010}}%
}]{%
seabold2010statsmodels}
\APACinsertmetastar {%
seabold2010statsmodels}%
\begin{APACrefauthors}%
Seabold, S.%
\BCBT {}\ \BBA {} Perktold, J.%
\end{APACrefauthors}%
\unskip\
\newblock
\APACrefYearMonthDay{2010}{}{}.
\newblock
\APACrefbtitle {statsmodels: {Econometric} and statistical modeling with
  {Python}.} {statsmodels: {Econometric} and statistical modeling with
  {Python}.}
\PrintBackRefs{\CurrentBib}

\bibitem [\protect \citeauthoryear {%
Trindade%
\ \BBA {} Macouin%
}{%
Trindade%
\ \BBA {} Macouin%
}{%
{\protect \APACyear {2007}}%
}]{%
Trindade-Macouin-2007:palaeolatitude}
\APACinsertmetastar {%
Trindade-Macouin-2007:palaeolatitude}%
\begin{APACrefauthors}%
Trindade, R\BPBI I\BPBI F.%
\BCBT {}\ \BBA {} Macouin, M.%
\end{APACrefauthors}%
\unskip\
\newblock
\APACrefYearMonthDay{2007}{}{}.
\newblock
{\BBOQ}\APACrefatitle {Palaeolatitude of glacial deposits and palaeogeography
  of {Neoproterozoic} ice ages} {Palaeolatitude of glacial deposits and
  palaeogeography of {Neoproterozoic} ice ages}.{\BBCQ}
\newblock
\APACjournalVolNumPages{Comptes Rendus Geoscience}{339}{3-4}{200--211}.
\PrintBackRefs{\CurrentBib}

\bibitem [\protect \citeauthoryear {%
Urey%
}{%
Urey%
}{%
{\protect \APACyear {1952}}%
}]{%
Urey-1952:early}
\APACinsertmetastar {%
Urey-1952:early}%
\begin{APACrefauthors}%
Urey, H\BPBI C.%
\end{APACrefauthors}%
\unskip\
\newblock
\APACrefYearMonthDay{1952}{}{}.
\newblock
{\BBOQ}\APACrefatitle {On the early chemical history of the earth and the
  origin of life} {On the early chemical history of the earth and the origin of
  life}.{\BBCQ}
\newblock
\APACjournalVolNumPages{Proceedings of the National Academy of
  Sciences}{38}{4}{351}.
\PrintBackRefs{\CurrentBib}

\bibitem [\protect \citeauthoryear {%
Walker%
, Hays%
\BCBL {}\ \BBA {} Kasting%
}{%
Walker%
\ \protect \BOthers {.}}{%
{\protect \APACyear {1981}}%
}]{%
Walker-Hays-Kasting-1981:negative}
\APACinsertmetastar {%
Walker-Hays-Kasting-1981:negative}%
\begin{APACrefauthors}%
Walker, J\BPBI C\BPBI G.%
, Hays, P\BPBI B.%
\BCBL {}\ \BBA {} Kasting, J\BPBI F.%
\end{APACrefauthors}%
\unskip\
\newblock
\APACrefYearMonthDay{1981}{}{}.
\newblock
{\BBOQ}\APACrefatitle {A negative feedback mechanism for the long-term
  stabilization of {Earth's} surface temperature} {A negative feedback
  mechanism for the long-term stabilization of {Earth's} surface
  temperature}.{\BBCQ}
\newblock
\APACjournalVolNumPages{Journal of Geophysical Research:
  Oceans}{86}{C10}{9776--9782}.
\PrintBackRefs{\CurrentBib}

\bibitem [\protect \citeauthoryear {%
Winnick%
\ \BBA {} Maher%
}{%
Winnick%
\ \BBA {} Maher%
}{%
{\protect \APACyear {2018}}%
}]{%
Winnick-Maher-2018:relationships}
\APACinsertmetastar {%
Winnick-Maher-2018:relationships}%
\begin{APACrefauthors}%
Winnick, M\BPBI J.%
\BCBT {}\ \BBA {} Maher, K.%
\end{APACrefauthors}%
\unskip\
\newblock
\APACrefYearMonthDay{2018}{}{}.
\newblock
{\BBOQ}\APACrefatitle {Relationships between {CO}$_2$, thermodynamic limits on
  silicate weathering, and the strength of the silicate weathering feedback}
  {Relationships between {CO}$_2$, thermodynamic limits on silicate weathering,
  and the strength of the silicate weathering feedback}.{\BBCQ}
\newblock
\APACjournalVolNumPages{Earth and Planetary Science Letters}{485}{}{111--120}.
\PrintBackRefs{\CurrentBib}

\end{thebibliography}
%




%
%
%
%
%

\end{document}